\documentclass[twocolumn,prb,showpacs,superscriptaddress]{revtex4}
\usepackage{graphicx}
\usepackage{dcolumn}
\usepackage{bm}

\begin{document}

\title{The Fulde-Ferrell-Larkin-Ovchinnikov states for the d-wave superconductor in the two-dimensional orthorhombic lattice}

\author{S. L. Liu}
\affiliation{College of Science, Nanjing
University of Posts and Telecommunications,\\
Nanjing 210003, China}

\author{Tao Zhou}
\affiliation{College of Science, Nanjing University of Aeronautics
and Astronautics, \\
Nanjing 210016, China}

\date{\today}

\begin{abstract}

The Fulde-Ferrell-Larkin-Ovchinnikov (FFLO) state of a
two-dimensional (2D) orthorhombic lattice superconductor
is studied based on the Bogoliubov-de-Gennes equations. It is
illustrated that the 2D FFLO state is suppressed
and only one-dimensional (1D) stripe state is stable. The stripe changes its orientation
with the increasing Zeeman field. There exists a crossover region where the gap structure has some local 2D features.
These results are
significantly different from those of the tetragonal lattice system.
 The local density
of states is also studied which can be checked and compared with experiments in future.
\end{abstract}


\maketitle

The Fulde-Ferrell-Larkin-Ovchinnikov (FFLO) state, known as the
finite momentum of the Copper pair, was predicted in the mid
of 1960s\cite{Fulde1964,Larkin1964}. It will occur when the Pauli
paramagnetism effect dominates over the orbital
effect\cite{Gruenberg1966}. Thus the low dimensional layered
superconducting (SC) materials with the magnetic field parallel to the SC layers are the strong candidates to
realize the FFLO state. Recently, signatures for possible FFLO
state have been found in various crystal systems, such as the
heavy fermion materials
$\mathrm{CeCoIn_{5}}$,\cite{Radovan2003,Bianchi2003,Kumagai2006}
the organic superconductors
$\mathrm{\lambda-(BETS)_{2}GaCl_{4}}$,\cite{Tanatar2002}
$\mathrm{\lambda-(BETS)_{2}FeCl_{4}}$\cite{Uji2001,Balicas2001},
$\mathrm{\kappa-(BEDT-TTF)_{2}Cu(NCS)_{2}}$,\cite{Shimahara1997,Manalo2000}
and the iron-based materials LiFeAs.\cite{Cho}
On the other hand, the realization of FFLO state in the cold atom
system is studied
intensively\cite{Loh2010,Koponen2007,Cai2010,Koponen2008,Parish2007}
and recently the experimental signatures are found in the
one-dimensional optical lattice.\cite{Liao2010}

The experimental development for the FFLO state in low dimensional
systems has attracted much attention. Theoretically the
calculation based on the lattice model is of great interest and
the results may compare with the experimental results of the
crystal systems or the cold atom system. One of the fundamental
issues of the FFLO state is the detailed gap structure, which can
be studied through minimizing the free energy\cite{Combescot2005},
or through the self-consistent calculation based on the
Eilenberger equation.\cite{Vorontsov2005,Mora2004} At the mean time, the
Bogoliubov-de-Gennes (BdG) technique has been a powerful tool to
study various imhomogenous SC states and it can also obtain the SC
order parameter self-consistently. Thus it is also an effective
tool to study the FFLO state especially in lattice system.
Previously, based on the BdG equations, the FFLO state was studied
intensively in the tetragonal lattice system. Many groups have
reported their results and numerically the gap structure
depends on the pairing
symmetry.\cite{Wang2006,Wang2007,Cui2008,Zuo2009,Zhou2009} For
$s$-wave pairing symmetry, the one-dimensional (1D) stripe-like
pattern was reported.\cite{Wang2006,Cui2008} For $d$-wave paring symmetry,
recently it was proposed that a transition from the 1D
stripe-like pattern to the two-dimensional (2D) checkerboard
pattern will occur as the exchange field increases.\cite{Zhou2009}
And the 2D pattern may change to the 1D pattern in presence of
the impurities.\cite{Wang2007,Zuo2009}

The above results of the gap structure for the tetragonal lattice
system are significantly different from the previous theoretical
results in the isotropic systems,\cite{Shimahara1998,Maki2002} as
discussed by Ref.[26], indicating that the
intrinsic symmetry of the crystal lattice should play an important
role in the gap structure of the FFLO states. For a real SC material,
there often exists a structural transition from tetragonal- to
orthorhombic-lattice with the variation of the doping level. Some
possible microscopic orders, such as the stripe
order\cite{Kivelson2003} or the nematic
order\cite{Hinkov2008,Daou2010,Doh2007,Vojta2010} may also lead to a
weak anisotropy in the ab-plane. Furthermore, the 1D FFLO stripe
in the tetragonal lattice itself may induce the in-plane
anisotropy. Therefore, the studies
of the FFLO states in an orthorhombic lattice system is of great interest
while so far
little attention is paid to the FFLO state on this system. 

In this paper, motivated by the above consideration, we calculate
the spatially distributed order parameter self-consistently based
on the BdG equations on a 2D lattice with $x-y$ anisotropic
hopping. Our results show that the 2D solution of the SC gap
structure will be suppressed and disappears as the orthorhombicity
strength $\gamma\equiv{t_{x}}/t_{y}$ increases. For the case of
$\gamma=1.05$, only stable 1D stripe state exists and it is also
found that the stripe changes its orientation with the increasing
Zeeman field. A crossover region exists between these two stripe states and some local 2D features are
obtained in this region. The local density of states (LDOS) is also studied
to distinguish the different states.

We start from the model on the 2D lattice with the
Zeeman splitting effect. The
model Hamiltonian can be written as,
\begin{eqnarray}\label{1}
    H=&&-\sum_{ij\sigma}(t_{ij}c_{i\sigma}^{\dagger}c_{j\sigma}+\mathrm{H.c.})-\sum_{i\sigma}(\mu+\sigma{h})c_{i\sigma}^{\dagger}c_{i\sigma}
    \nonumber\\
    &&+\sum_{ij}(\Delta_{ij}c_{i\uparrow}^{\dagger}c_{j\downarrow}^{\dagger}+\mathrm{H.c.}),
\end{eqnarray}
where $t_{ij}$ are the hopping constants and $\mu$ is the chemical
potential. $\sigma{h}$ is the Zeeman
energy term, caused by the in-plane magnetic field, with $\sigma=\pm1$ representing the spin-up and
spin-down electrons respectively.

This Hamiltonian can be diagonalized by solving the BdG equations,
\begin{equation}
\sum_j \left( \begin{array}{cc}
 H_{i j} & \Delta_{i j}  \\
 \Delta^{*}_{i j} & -H^{*}_{ij}
\end{array}
\right) \left( \begin{array}{c}
u^{n}_{j\uparrow}\\v^{n}_{j\downarrow}
\end{array}
\right) =E_n \left( \begin{array}{c}
u^{n}_{i\uparrow}\\v^{n}_{i{\downarrow}}
\end{array}
\right),\label{2}
\end{equation}
where $H_{ij}$ is expressed by
\begin{equation}\label{3}
     H_{ij}=-t_{ij}-(\mu+\sigma{h})\delta_{ij}.
\end{equation}
The SC order parameter and the local electron density $n_{i}$
are obtained self-consistently:
\begin{equation}\label{4}
    \Delta_{ij}=\frac{V_{ij}}{4}\sum_{n}(u_{i\uparrow}^{n}\upsilon_{j\downarrow}^{n*}+u_{j\uparrow}^{n}\upsilon_{i\downarrow}^{n*})\mathrm{tanh}\bigg(\frac{E_{n}}{2k_{B}T}\bigg),
\end{equation}
\begin{equation}\label{5}
    n_{i}=\sum_{n}|u_{i\uparrow}^{n}|^{2}f(E_{n})+\sum_{n}|\upsilon_{i\downarrow}^{n}|^{2}[1-f(E_{n})].
\end{equation}
Here
$f(x)$ is the Fermi distribution function.
$V_{ij}$ is the pairing strength. In the present work, we consider the nearest neighbor (NN) pairing with $V_{ij}=V\delta_{i,j\pm\hat{\alpha}}$.
For the tetragonal lattice system the NN pairing will reproduce the $d_{x^2-y^2}$-pairing symmetry. The $d$-wave order parameter at the site $i$ can be defined as $\Delta^d_{i}=1/4(\Delta_{i,i+\hat{x}}+\Delta_{i,i-\hat{x}}-\Delta_{i,i+\hat{y}}-\Delta_{i,i-\hat{y}})$.
For the case of orthorhombic lattice the four-fold symmetry is broken so that a $s$-wave component will be induced and is expected to increase as the orthorhombicity strength $\gamma$ increases. The $s$-wave component is defined as $\Delta^s_{i}=1/4(\Delta_{i,i+\hat{x}}+\Delta_{i,i-\hat{x}}+\Delta_{i,i+\hat{y}}+\Delta_{i,i-\hat{y}})$. Since we consider only weak $x-y$ isotropy in the present work with $\gamma \leq 1.05$. Our numerical results show that the $s$-wave component is quite small [$\Delta^{s}/\Delta^{d}<0.03$]. Thus in our following presented results we neglect the small $s$-wave component and use the above $d$-wave order parameter as the definition of the on-site order parameter. We also define the The magnetization $m_{i}$ as
$m_{i}=\langle{S^{z}_i}\rangle=\hbar/2\langle{n_{i\uparrow}}-n_{i\downarrow}\rangle$.

The LDOS is expressed by
\begin{equation}\label{6}
    \rho_{i}(\omega)=\sum_{n}[|u_{i\uparrow}^{n}|^{2}\delta(E_{n}-\omega)+|\upsilon_{i\downarrow}^{n}|^{2}\delta(E_{n}+\omega)],
\end{equation}
where the delta function $\delta(x)$ has been approximated by
$\Gamma/\pi(x^{2}+\Gamma^{2})$ with $\Gamma=0.02$.

In the following calculation, we consider the NN hopping with the hopping constant in $x$
direction being $t_{x}=1.0$ and that in $y$ direction
$t_{y}=t_{x}/\gamma$. 
Here, $\gamma$ represents the orthorhombicity strength.
The pairing potential $V$ and the filling electron density $n$ are
chosen as $V=1.3$ and $n=0.84$ (hole-doped samples with doping
$\delta=0.16$). The calculation is made on a $48\times48$ lattice
with the periodic boundary condition, and the randomly distributed
initial values of the order parameters are chosen. The
$10\times10$ supercell is used to calculate the LDOS.

\begin{figure}
\begin{center}
\includegraphics{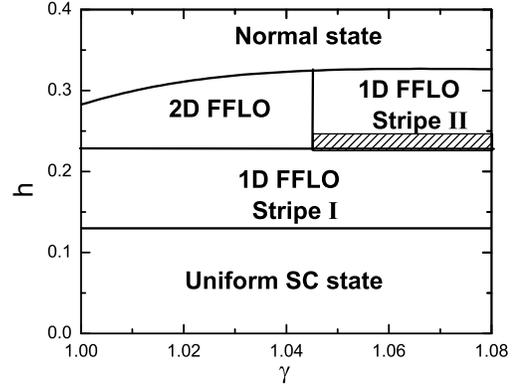}
\caption {Phase diagram of an 2D orthorhombic
lattice superconductor in the parallel magnetic field.
$h=g\mu_{B}H$ is the exchange field and $\gamma$ the orthorhombicity strength. The crossover region is indicated by the shadow.}
\end{center}
\end{figure}

Our main results are summarized in Fig. 1, where the phase diagram
is plotted. As one can see, for low anisotropy, the whole SC state is divided into three regions,
namely, the uniform SC state, the 1D FFLO state, and the 2D FFLO
state, which is consistent with the previous
results.\cite{Zhou2009} For $\gamma=1.05$ and at low fields, the SC
state still includes the uniform SC state and the 1D FFLO
state. At $0.22\leq{h}<0.23$, as indicated by the shadow, there exists a crossover region. In this region, the gap structure
forms the coexistence of the 1D stripe-like pattern and some local 2D features.
As the exchange field increases further, as seen,
another 1D FFLO state shows up. The orientation of the stripe pattern is different from that of the low field, namely,
for the lower field FFLO state the stripe 
is parallel to the $x$ direction (stripe I) and that of the
higher field one is parallel to the $y$ direction (stripe II).
We also
note that the upper critical field $H_{c2}$ increases with the
increasing $\gamma$.

We now studied the self-consistent results of the order parameter and magnetization 
with $\gamma=1.05$ in Fig.2.
The order parameters are plotted in Figs.2(a)-2(e). As seen, for the weaker magnetic field, the order parameter
is uniform [Fig. 2(a)]. The SC order forms the stripe
pattern as the exchange field increases to 0.2, as seen in Fig.2(b). The periodicity of the order parameter is about 24 along
$y$ direction as $h=0.20$, which is consistent with the previous
results.\cite{Zhou2009} For the case of $h=0.22$ and $h=0.225$, a crossover
region from stripe I state to stripe II state occurs, indicated in
Figs. 2(c) and 2(d). As seen in Fig. 2(c), the gap structure has some 2D features around the point 'A'
with the stripe II state showing up gradually.
For $h=0.225$ [Fig. 2(d)], the stripe II state dominates over the stripe I one and the
2D FFLO features are almost suppressed. As $h$ increases
further, i.e. for $h=0.23$, the SC order forms the stripe pattern
again with the stripe parallelling to the $y$ axis.

\begin{figure}
    \includegraphics[width=1.5in]{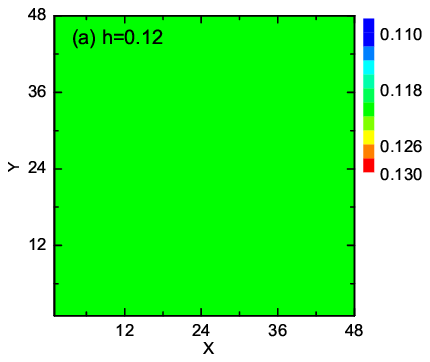}
    \includegraphics[width=1.5in]{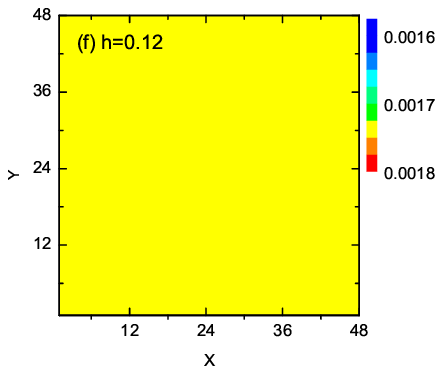}
    \includegraphics[width=1.5in]{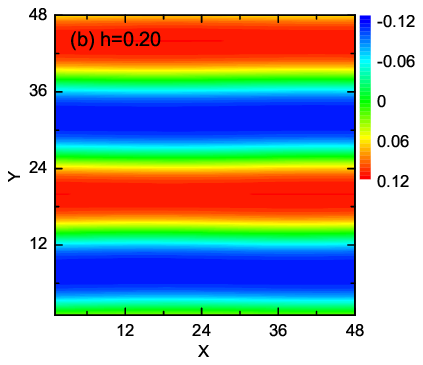}
    \includegraphics[width=1.5in]{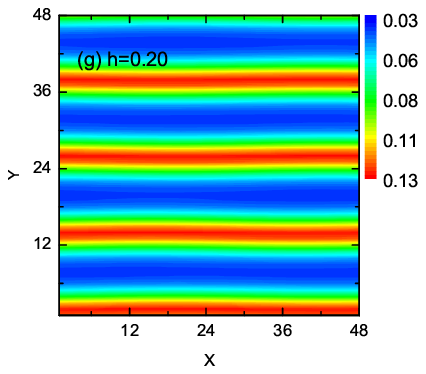}
    \includegraphics[width=1.5in]{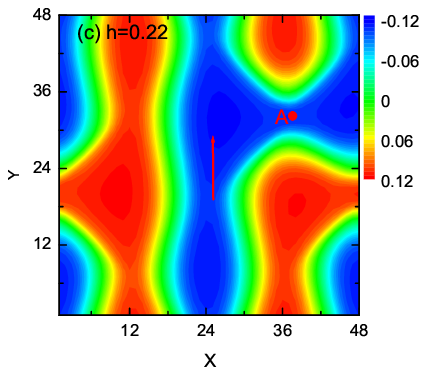}
    \includegraphics[width=1.5in]{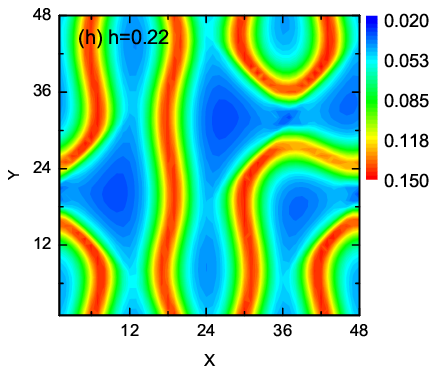}
    \includegraphics[width=1.5in]{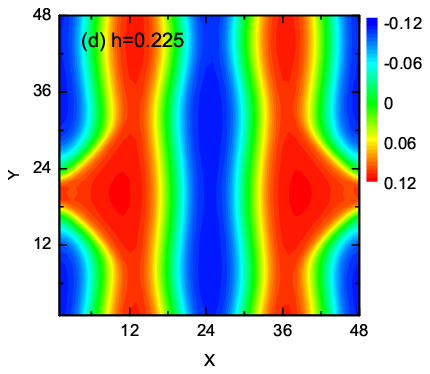}
    \includegraphics[width=1.5in]{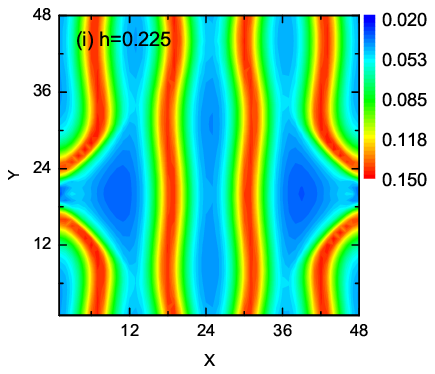}
    \includegraphics[width=1.5in]{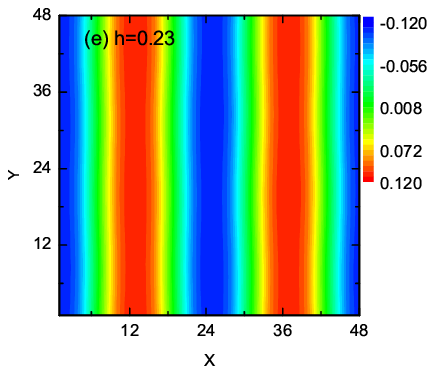}
    \includegraphics[width=1.5in]{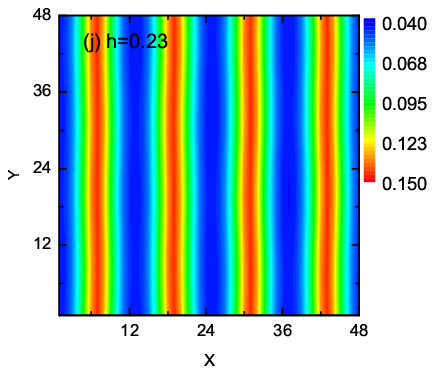}
\caption {(Color online) Plots of the order parameter $\Delta$ and and
the magnetization $m_{i}$ as a function of position for various
Zeeman fields $h$ with $T=10^{-5}$ at anisotropic hopping
parameter $\gamma=1.05$.}
\end{figure}

The spatial distributions of the magnetization (with the units
$\hbar$) are shown in Figs. 2(f)-2(j). As seen in Fig. 2(f), in
the uniform phase, the distribution is also uniform, where the
magnetization is very weak (about 0.0017) due to the suppression
by the SC order. In the 1D FFLO stripe
I state, as seen in Fig. 2(g), the pattern also forms 1D stripe
but the periodicity is only half of that of the order parameter.
The intensity is largest along the nodal lines and is
suppressed when the SC order parameter increases. It reaches the
minimum value as the SC order is maximum. These features are similar to previous results 
in tetragonal lattice system~\cite{Zhou2009}. In the crossover region,
such as $h=0.22$, the magnetization also behaves the coexistence
of the 1D and 2D character. As $h$ increasing further, the magnetization forms
1D stripe again with the orientation 
changing from $y$ axis to $x$ axis.

\begin{figure}
\begin{center}
\includegraphics[width=2.4in]{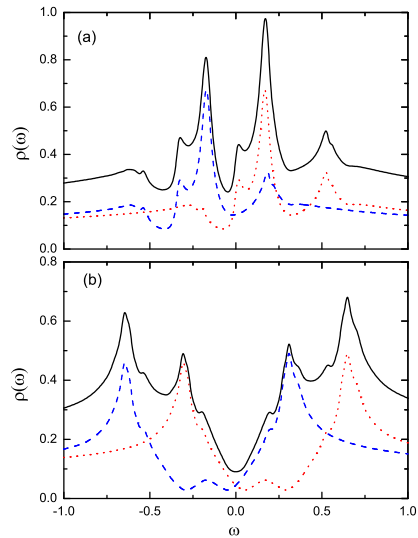}
\caption {(Color online) The LDOS spectra for 1D stripe I state at $h=0.17$.
Panel (a) and (b) are the LDOS spectra at the nodal line and at
the site where the order parameter is maximum, respectively. The
(blue) dashed and (red) dotted lines, and the (black) solid line are spin-up
LDOS, spin-down LDOS, and whole LDOS, respectively.}
\end{center}
\end{figure}

Now let us study the LDOS spectra.
The LDOS from Eq.6 includes two terms, i.e., the spin-up LDOS and spin-down LDOS, respectively.
In presence of
the Zeeman field, due to the Pauli paramagnetic effect, the spin-up LDOS and spin-down one sperate with the spin-up one
 shifting to the left and the
spin-down one to the right, respectively. The calculated LDOS spectra in
the uniform phase are similar to the previous
report,\cite{Zhou2009} which is not shown here. The LDOS spectra in the 1D stripe I phase with $h=0.17$ are shown
in Fig. 3. Fig. 3(a) is for the site on the nodal line. As seen,
the spin-up LDOS spectra show two low-energy peaks locating at
$\omega=-0.17$ and $-0.32$. The SC coherent peak is suppressed.
The spin-down LDOS shifts to the right with the two low-energy
peaks at $\omega=0.02$ and $0.17$. These in-gap peaks originate
from the finite energy Andreev bound states due to the sign change
in the order parameter across the nodal lines, which is the
signature of 1D FFLO state, as also discussed previously~\cite{Wang2006,Zhou2009}. The peaks at the negative energy in
total LDOS comes from the spin-up LDOS and the peaks at the
positive energy are contributed by the spin-down LDOS. The
intensity of the in-gap peaks will decrease as the site moves away
from the nodal line. As seen from Fig. 3(b), at the site
where the order parameter is maximum, the in-gap peaks are turned
to be a hump at the midgap position for both spin-up and spin-down
LDOS spectra. The SC coherent peaks are seen clearly. The midgap
hump is so weak that it is concealed in the whole LDOS spectrum.

\begin{figure}
\begin{center}
\includegraphics{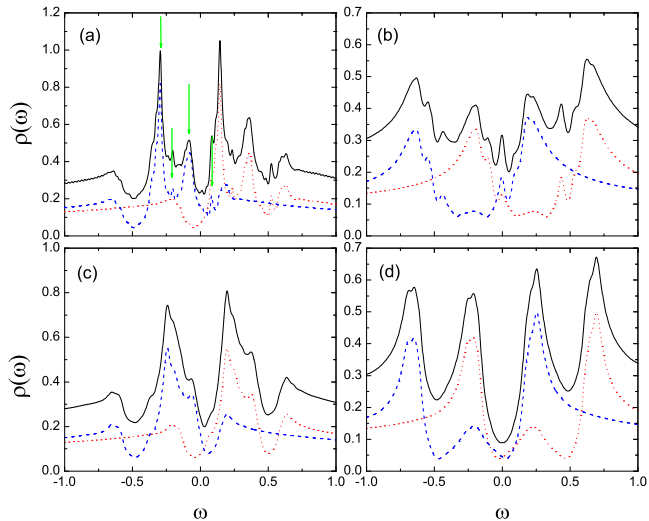}
\caption {(Color online) The LDOS spectra in the crossover region with $h=0.22$. Panel
(a) is the spectrum at the saddle ponit where two nodal lines
intersect. Panel (b) is the spectrum at the site where the order
parameter is maximum in the 2D FFLO region. Panels (c) and (d) are
the spectra at the nodal line and at the site where the order
parameter is maximum in the 1D FFLO region, respectively. The
(blue) dashed and (red) dotted lines, and the (black) solid line are spin-up
LDOS, spin-down LDOS, and whole LDOS, respectively. }
\end{center}
\end{figure}

The LDOS spectra for the crossover region with $h=0.22$ are presented in Fig.4.
 The spectrum at the saddle point in the local 2D FFLO
region [near the point "A" in Fig. 2(c)] is presented in Fig.
4(a). Although the gap structure is quite different from previous results in tetragonal
 lattice system.
While actually for this case the spectrum is similar to that of the saddle point's spectrum in the 2D checkerboard FFLO state~\cite{Wang2006,Zhou2009}, namely, two kinds of Andreev bound states exist.
 As a result, four in-gap peaks exist in the spin-up LDOS spectra at the
energies -0.300, -0.208, -0.052, and 0.081 (indicated by the
arrows). At the site where the order parameter is maximum, the
LDOS spectrum [Fig. 4(b)] is very complicated, while the two
coherence peaks outside can be seen clearly. Shown in Fig. 4(c) is
the LDOS spectrum on the nodal line in the 1D FFLO region. There
are also two in-gap peaks in both the spin-up and spin-down LDOS
spectra, indicating the 1D characteristics. The SC coherent peaks
are suppressed again, similar to the case of 1D stripe phase [Fig.
3(a)]. At the site where the order parameter is maximum [Fig.
4(d)], the in-gap peaks are turned to be a hump at the midgap
position for both the spin-up and spin-down LDOS spectra, while
the SC coherence peaks are clearly seen, which leads to the four
peaks in the whole LDOS spectrum. Summing up the above results, it
is found that in the crossover region, the FFLO indeed includes
both the 1D and 2D FFLO characters, which is
consistent with the spatial distribution of the order parameter
and the magnetization (Fig. 2).

\begin{figure}
\begin{center}
\includegraphics[width=2.4in]{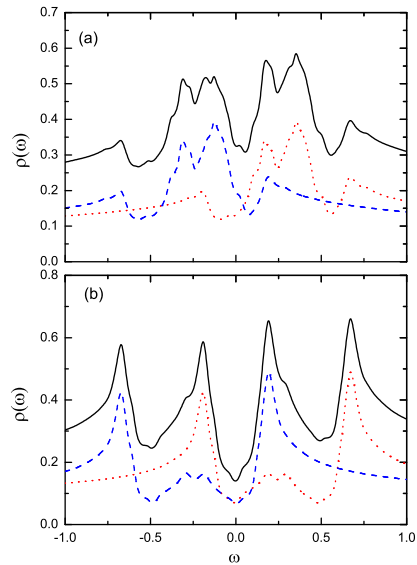}
\caption {(Color online) Similar to Fig. 3, but for
h=0.24.}
\end{center}
\end{figure}

The LDOS spectra of the 1D stripe II state with $h=0.24$ is
presented in Fig. 5. The result at the nodal line is shown in Fig.
5(a). It can be seen that there are two in-gap peaks of both the
spin-up and spin-down LDOS, while the SC coherence peaks are
suppressed. Presented in Fig. 5(b) is the LDOS spectrum at the
site where the order parameter is maximum. It is found that the
in-gap peaks are almost suppressed completely, while the SC
coherence peaks are clearly seen. Thus, the feature of the LDOS
spectra of the 1D stripe II state is similar to that of the 1D
stripe I state.

We have shown the LDOS spectra of the three different phases,
namely the 1D strip 1 phase, the crossover region, and the 1D stripe II
phase. As seen in Fig. 4(a)-4(d), the spectra of the Q2D state are
quite different from that of the other two 1D stripe phases. In
the Q2D state, the 1D and 2D FFLO states coexist in different
regions. These features are expected to be detected by the scanning tunneling microscopy (STM) experiments.

\begin{figure}
    \includegraphics[width=1.5in]{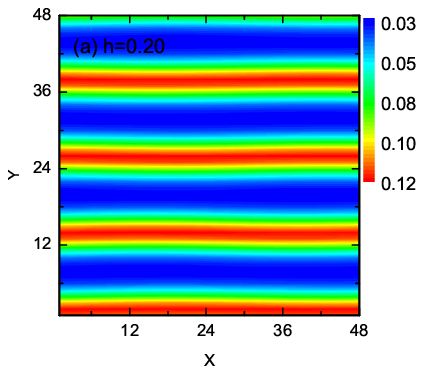}
    \includegraphics[width=1.5in]{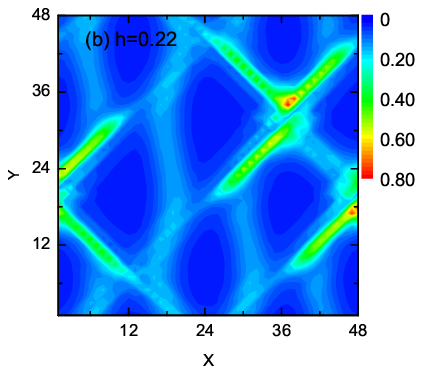}
    \includegraphics[width=1.5in]{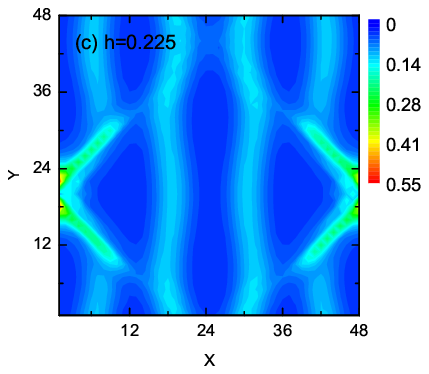}
    \includegraphics[width=1.5in]{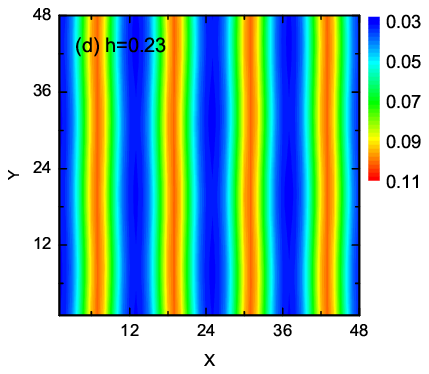}
\caption {(Color online) The LDOS maps for quasiparticles with $\omega=0$ and different exchange fields. }
\end{figure}

To distinguish the two stripe phases, we present the LDOS images
with $\omega=0$. Shown in Fig. 6(a) is the LDOS image at $h=0.20$
in the 1D stripe I state. As seen, in this state it also forms 1D
stripe-like pattern similar to the spatial distribution of the
order parameter but the periodicity is only half of that of the
order parameter. The intensity is largest along the nodal
lines and is suppressed when the SC order parameter increases. It
reaches the minimum value as the SC order is maximum. The stripe
parallels to the $x$ axis, consistent with the gap structure. The
images in the crossover region are presented in Figs. 6(b) and 6(c)
with $h=0.22$ and $h=0.225$, respectively. As seen from Fig. 6(b),
the LDOS image in this state also behaves the Q2D characteristics,
i.e., the stripe-like and the local 2D patterns
coexist. The intensity is largest at the saddle points where
two nodal lines intersect. It reaches the minimum value as the SC
order is maximum. The LDOS image at $h=0.23$ in the 1D stripe II
phase is presented in Fig. 6(d). It is found that the image forms
stripe-like pattern again with the orientation switching to the $y$ axis.
Therefore, it can be used to distinguish the two stripe phases by
the STM experiments.

In summary, based on the BdG equations and the $d$-wave superconductivity, the phase
diagram and the order-parameter structure is studied in 
presence of the external exchange field of a 2D orthorhombic
lattice superconductor. It is found that for weak
anisotropic hoping, i.e. $\gamma=1.05$, the 2D FFLO state is
suppressed and only the 1D state exists. The orientation of the
stripe changes as the exchange field increases. The local Q2D
state exists in the crossover region. These results are much
different from those of the tetragonal lattice system. Our result
suggests that the symmetry of the lattice plays an important role
in the gap structure of the FFLO state. The local Q2D state can be
detected by the STM experiments according to the calculated
spectra of LDOS. The LDOS images with $\omega=0$ in these phases
are also presented, which can be used to distinguish the 
different phases.

We thank Dr. Chen Wei and Dr. Xue Hongtao for the
technical support. This work was supported by the Scientific Research Foundation from Nanjing University of Aeronautics and Astronautics.


\begin{thebibliography}{99}

\bibitem{Fulde1964}P. Fulde and R. A. Ferrell, Phys. Rev. {\bf 135}, A550 (1964).

\bibitem{Larkin1964}A. I. Larkin and Yu. N. Ovchinnikov, Zh. Eksp. Teor. Fiz. {\bf 47}, 1136
(1964) [Sov. Phys. JETP {\bf 20}, 762 (1965)].

\bibitem{Gruenberg1966} L. W. Gruenberg and L. Gunther, Phys. Rev. Lett. {\bf 16}, 996 (1966).

\bibitem{Radovan2003}H. A. Radovan, N. A. Fortune, T. P. Murphy, S. T. Hannahs, E. C.
Palm, S. W. Tozer, and D. Hall, Nature (London) {\bf 425}, 51
(2003).

\bibitem{Bianchi2003}A. Bianchi, R. Movshovich, C. Capan, P. G. Pagliuso, and J. L.
Sarrao, Phys. Rev. Lett. {\bf 91}, 187004 (2003).

\bibitem{Kumagai2006}K. Kumagai, M. Saitoh, T. Oyaizu, Y. Furukawa, S. Takashima, M.
Nohara, H. Takagi, and Y. Matsuda, Phys. Rev. Lett. {\bf 97},
227002 (2006).

\bibitem{Tanatar2002}M. A. Tanatar, T. Ishiguro, H. Tanaka, and H. Kobayashi, Phys.
Rev. B {\bf 66}, 134503 (2002).

\bibitem{Uji2001} S. Uji, H. Shinagawa, T. Terashima, T. Yakabe, Y. Terai, M. Tokumoto, A. Kobayashi, H.
Tanaka, and H. Kobayashi, Nature (London) {\bf 410}, 908 (2001).

\bibitem{Balicas2001} L. Balicas, J. S. Brooks, K. Storr, S. Uji, M. Tokumoto, H.
Tanaka, H. Kobayashi, A. Kobayashi, V. Barzykin, and L. P. Gorkov,
Phys. Rev. Lett. {\bf 87}, 067002 (2001).

\bibitem{Shimahara1997}H. Shimahara, J. Phys. Soc. Jpn. {\bf 66}, 541 (1997).

\bibitem{Manalo2000} S. Manalo and U. Klein, J. Phys.: Condens. Matter {\bf 12}, L471
(2000).
\bibitem{Cho} K. Cho, H. Kim, M. A. Tanatar, Y. J. Song, Y. S. Kwon,
W. A. Coniglio, C. C. Agosta, A. Gurevich, and R. Prozorov, Phys. Rev. B {\bf 83}, R060502 (2011).

\bibitem{Cai2010}Zi Cai, Yupeng Wang, and Congjun Wu,
arXiv:1009.3257.

\bibitem{Loh2010}Y. L. Loh and N. Trivedi, Phys. Rev. Lett. {\bf 104}, 165302
(2010).

\bibitem{Koponen2007}T. K. Koponen, T. Paananen, J. P. Martikainen, and P. T\"{o}rm\"{a},
Phys. Rev. Lett. {\bf 99}, 120403 (2007).

\bibitem{Koponen2008}T. K. Koponen, T. Paananen, J. P. Martikainen, M. R. Bakhtiari and
P. T\"{o}rm\"{a},, N. J. Phys. {\bf 10}, 045014 (2008).

\bibitem{Parish2007}M. M. Parish, S. K. Baur, E. J. Mueller, and D. A. Huse, Phys.
Rev. Lett. {\bf 99}, 250403 (2007).

\bibitem{Liao2010} Yean-an Liao, Ann Sophie C. Rittner, Tobias Paprotta, Wenhui Li, Guthrie B.
Partridge, Randall G. Hulet, Stefan K. Baur and Erich J. Mueller,
Nature (London) {\bf 467} 567 (2010).

\bibitem{Combescot2005}R. Combescot and C. Mora, Phys. Rev. B {\bf 71}, 144517 (2005).

\bibitem{Vorontsov2005}A. B. Vorontsov, J. A. Sauls, and M. J. Graf, Phys. Rev. B {\bf
72}, 184501 {2005}.

\bibitem{Mora2004}C. Mora and R. Combescot, Europhys. Lett. {\bf 66}, 833 (2004).

\bibitem{Wang2006}Qian Wang, H.-Y. Chen, C.-R. Hu, and C. S. Ting, Phys. Rev.
Lett. {\bf 96}, 117006 (2006).

\bibitem{Cui2008}Q. Cui and K. Yang, Phys. Rev. B {\bf 78}, 054501 (2008).

\bibitem{Wang2007}Q. Wang, C.-R. Hu, and C.-S. Ting, Phys. Rev. B {\bf 75}, 184515 (2007).

\bibitem{Zuo2009}X.-J. Zuo and C.-D. Gong, EPL {\bf 86}, 47004 (2009).

\bibitem{Zhou2009}Tao Zhou and C. S. Ting, Phys. Rev. B {\bf 80},
224515 (2009).

\bibitem{Shimahara1998}H. Shimahara, J. Phys. Soc. Jpn. {\bf 67}, 736 (1998).

\bibitem{Maki2002}K. Maki and H. Won, Physica B (Amsterdam) {\bf 322}, 315 (2002).

\bibitem{Kivelson2003}S. A. Kivelson, I. P. Bindloss, E. Fradkin, V. Oganesyan, J. M.
Tranquada, A. Kapitulnik, and C. Howald, Rev. Mod. Phys. {\bf 75},
1201 (2003).

\bibitem{Hinkov2008}V. Hinkov, D. Haug, B. Fauqu¡äe, P. Bourges, Y. Sidis, A. Ivanov,
C. Bernhard, C. T. Lin, and B. Keimer, Science {\bf 319}, 597
(2008).

\bibitem{Daou2010}R. Daou, J. Chang, D. LeBoeuf, O. Cyr-Choiniere, F. Laliberte, N.
Doiron-Leyraud, B. J. Ramshaw, R. Liang, D. A. Bonn, W. N. Hardy,
and L. Taillefer, Nature {\bf 463}, 519 (2010).

\bibitem{Vojta2010}Matthias Vojta, Eur. Phys. J. Special Topics {\bf 188}, 49 (2010).

\bibitem{Doh2007}Hyeonjin Doh and Hae-Young Kee, Phys. Rev. B {\bf 75}, 233102
 (2007).







\end{thebibliography}
\end{document}